\documentclass[fleqn,usenatbib,useAMS]{mnras}

\usepackage{graphicx}	
\usepackage{amsmath}	
\usepackage{amssymb}	
\usepackage{multicol}        
\usepackage{bm}		
\usepackage{pdflscape}	

\usepackage{multirow}

\usepackage[T1]{fontenc}
\usepackage{ae,aecompl}

\usepackage{newtxtext,newtxmath}

\title{Sensitivity tests of cosmic velocity fields to massive neutrinos}

\author[Shuren Zhou et al.]{Shuren Zhou$^{1,2}$, Zhenjie Liu$^{3,1}$, Qinglin Ma$^{4,1}$, Yu Liu$^{3}$, Le Zhang$^{1,2}$\thanks{zhangle7@mail.sysu.edu.cn}, Xiao-Dong Li$^{1,2}$\thanks{lixiaod25@mail.sysu.edu.cn}, Yang Wang$^{5,2}$\thanks{wangy18@pcl.ac.cn},
\newauthor
Xin Wang$^{1,2}\thanks{wangxin35@mail.sysu.edu.cn}$,Yu Yu$^{3}$, Hao-Ran Yu$^{6}$, Yi Zheng$^{1,2}$\thanks{zhengyi27@mail.sysu.edu.cn} \\
$^{1}$School of Physics and Astronomy, Sun Yat-Sen University, Guangzhou, 510297, P. R. China\\
$^{2}$CSST Science Center for the Guangdong-Hong kong-Macau Greater Bay Area, SYSU\\
$^{3}$Department of Astronomy, Shanghai Jiao Tong University, Shanghai, 200240, P. R. China\\
$^{4}$Department of Astronomy, Tsinghua University, Beijing, 100084, P.R. China\\
$^{5}$Department of Mathematics and Theories, Peng Cheng Laboratory, No.2, Xingke 1st Street, Shenzhen, 518000, P. R. China\\
$^{6}$Department of Astronomy, Xiamen University, Xiamen, Fujian, 361005, P.R. China
}
\date{Accepted XXX. Received YYY; in original form ZZZ}

\pubyear{2021}

\begin{document}
\label{firstpage}
\pagerange{\pageref{firstpage}--\pageref{lastpage}}
\maketitle

\begin{abstract}
We investigate impacts of massive neutrinos on the cosmic velocity fields, employing high-resolution cosmological $N$-body simulations provided by the information-optimized $\texttt{CUBE}$ code, where cosmic neutrinos are evolved using collisionless hydrodynamics and their perturbations can be accurately resolved. In this study we focus, for the first time, on the analysis of massive-neutrino induced suppression effects in various cosmic velocity field components of velocity magnitude, divergence, vorticity and dispersion. By varying the neutrino mass sum $M_\nu$ from 0 -- 0.4 eV, the simulations show that, the power spectra of vorticity  --- exclusively sourced by non-linear structure formation that is affected by massive neutrinos significantly ---  is very sensitive to the mass sum, which potentially provide novel signatures in detecting massive neutrinos. Furthermore, using the chi-square statistic, we quantitatively test the sensitivity of the density and velocity power spectra to the neutrino mass sum. Indeed, we find that, the vorticity spectrum has the highest sensitivity, and the null hypothesis of massless neutrinos is incompatible with both vorticity and divergence spectra from $M_\nu=0.1$ eV at high significance ($p$-value $= 0.03$ and $0.07$, respectively). These results demonstrate clearly the importance of peculiar velocity field measurements, in particular of vorticity and divergence components, in determination of neutrino mass and mass hierarchy.
\end{abstract}





\section{Introduction}\label{s1}

Neutrinos are one of the most mysterious particles observed in nature: they are only weakly interacting with all the other elementary particles in the Standard Model of particle physics, but their masses are several orders of magnitude smaller than those of leptons and quarks. In spite of decades of numerous experimental efforts, the determination of their masses remains elusive. The discovery of neutrino oscillations~\citep{PhysRevD.46.3720,PhysRevLett.81.1158,PhysRevLett.92.181301} between the flavor eigenstates has conclusively revealed that neutrinos are not massless, and currently, only the differences between the squared masses of the three neutrino species are measured~\citep{Olive_2016}: $\Delta m^2_{21} \equiv m^2_2 -m^2_1 \approx 7.54^{+0.26}_{-0.22} \times 10^{-5} {\rm eV^2}$ and $| \Delta m^2_{31}| \equiv |m^2_3 - m^2_1| \approx 2.46^{+0.06}_{-0.06} \times 10^{-3} {\rm eV^2}$. However, it is difficult to determine the total neutrino mass by using particle physics experiments, e.g.,~\citep{WOLF2010442,Esfahani_2017}, as they are only sensitive to the lightest neutrino mass. Upcoming laboratory-based experiments, such as tritium endpoint and double beta decay experiments, are promising to improve bounds on the neutrino mass scale (\cite{Drexlin_2013} for review). From the current oscillation data, since the sign of $\Delta m^2_{31}$ is unknown, there are two possible hierarchies of neutrino masses: $m_1 < m_2 \ll m_3$ (normal hierarchy for positive $\Delta m^2_{31}$) and $m_3 \ll m_1 < m_2$ (inverted hierarchy for negative one), with minimum neutrino mass sums of $M_\nu\equiv \sum m_i = 0.06$ eV and 0.1 eV, respectively.

In cosmology, effects of massive neutrinos on cosmological observables, such as
Cosmic Microwave Background (CMB) and Large-Scale Structure (LSS), have been extensively investigated in the literature~\citep{LESGOURGUES_2006,Wong_2011,2012arXiv1212.6154L,Lesgourgues_2014,Abazajian_2015,Archidiacono_2017} and these effects offer a promising independent probe of neutrino masses beyond particle physics experiments. At early times, neutrinos with masses $\ll$ eV remain relativistic and behave like radiation at the time of photon decoupling, so that the influences of their mass on the photon perturbation and its evolution are very limited. Hence, massive neutrinos can only affect the background evolution for the primary CMB anisotropies and Integrated Sachs-Wolfe (ISW) effect for the secondary anisotropies. However, these effects are potentially degenerate with other cosmological parameters. At late times, neutrinos substantially influence the evolution of matter perturbations. Neutrinos with small masses have large thermal velocities and do not cluster below their free-streaming scale (around $110$ Mpc today for $M_\nu=0.1$ eV), which can cause a suppression of small-scale matter power spectrum. On larger scales, they cluster in the same way just as cold dark matter and baryonic matter. Currently, a combination of the CMB and baryonic acoustic oscillation measurements places the tightest constraint on the upper bound of neutrino mass sum, $M_\nu<0.12$ eV (2-$\sigma$) for a flat $\Lambda$CDM cosmology~\citep{2020A&A...641A...6P}, which is, however, not sensitive enough to discriminate these two hierarchies at high significance. Other cosmological observables such as weak lensing~\citep{2011JCAP...06..027V, 2019JCAP...06..019M, 2019PhRvD..99f3527L, 2019JCAP...05..043C}, Lyman-$\alpha$ forest~\citep{2016JCAP...11..015B,2015JCAP...11..011P,2010JCAP...06..015V} can also provide independent and complementary constraints on the mass sum. Also, the mass bound depends on underlying cosmological models and different cosmological data adopted~\citep{2020JCAP...07..037C, 2020IJMPD..2950088L, 2020SCPMA..6380411Z,2019PhRvD..99f3527L}.

The massive-neutrino induced effects on commonly used observables for two-point statistics have been well studied, such as two-point correlation functions~\citep{2015JCAP...11..018M, 2011MNRAS.418..346M, 2019MNRAS.488.4413K, 2014JCAP...03..011V}, halo mass functions~\citep{2010JCAP...09..014B, 2018JCAP...03..049L, 2011MNRAS.418..346M, 2014JCAP...02..049C, 2013JCAP...12..012C}, the matter power spectrum ~\citep{2011MNRAS.410.1647A, 2020PhRvD.101f3515L, 2018JCAP...03..049L, 2010JCAP...06..015V, 2012arXiv1212.6154L, 2012MNRAS.420.2551B}.

However, those observables can not well capture the full information content in the nonlinear regime where the neutrinos are expected to play an important role. Recently, there has been a growing interest in statistics beyond two-point correlation in order to fully characterize massive-neutrino effects encoded in the higher-order information of data, such as by using bispectrum~\citep{2019JCAP...05..043C,2018JCAP...03..003R}, Minkowski functionals~\citep{2019JCAP...06..019M, 2020PhRvD.101f3515L}, lensing peak counts~\citep{2019PhRvD..99f3527L}, the marked power spectrum~\citep{Massara_2021}, which have been shown promisingly to tighten the limits on neutrino masses.

Beyond the density field, the cosmological velocity field of cold dark matter actually contains valuable information of the non-linear evolution of the late-time universe~\citep{2009PhRvD..80d3504P}, e.g.,  massive neutrinos would leave imprints on the first and second moment of the two-point relative velocities~\citep{2020A&A...644A.170K}. The velocity field  would become very sensitive to the neutrino masses and may offer new opportunities to detect the signature of neutrinos, in the sense that massive neutrinos suppress the structure growth and thus notably affect the non-linear structure formation and gravitational collapse that source velocity vorticity and anisotropic velocity dispersion. Therefore, in this study, we use the $N$-body simulations of $\texttt{CUBE}$~\citep{2018ApJS..237...24Y} to accurately test the sensitivity of various velocity components to the total neutrino mass, by measuring the power spectra of velocity magnitude, divergence, vorticity and dispersion.  

This paper is organized as follows. We first introduce the $N$-body simulations used in this study and give a brief description of two-point statistics of different velocity components in Sect.~\ref{s2}. Next, in Sects.~\ref{s3} and ~\ref{s4}, we present effects from massive neutrinos in the CDM density field and different velocity components estimated from applying the Delaunay method to our $N$-body simulations. In Sect.~\ref{s5}, we quantify the sensitivity of those fields to massive neutrinos and we draw our conclusions in Sect.~\ref{s6}.

\section{Method and simulations}\label{s2}

\subsection{Method}
Here we will describe the numerical implementation of various velocity components in the $N$-body code $\texttt{CUBE}$. The velocity field, $\bm{v}$, as any vector field, can be split into gradient and rotational parts, and is completely described by its divergence, $\theta \equiv \nabla \cdot \bm{v} $ and its vorticity, $\bm{\omega} = \nabla \times \bm{v}$, which, in Fourier space, become purely radial and transversal velocity modes, respectively, defined by  $\theta(\textbf{k}) = i \textbf{k} \cdot \bm{v}(\textbf{k})$ and ${\bm \omega}(\textbf{k}) = i \textbf{k} \times \bm{v}(\textbf{k})$. In a spatially homogeneous Universe, the power spectra of the velocity, divergence, vorticity and velocity magnitude as well as the overdensity field are given by 
\begin{align}
\langle \theta (\textbf{k}) \theta^* (\textbf{k}^\prime) \rangle =& (2\pi)^3 P_{\theta \theta}(k) \delta ( \textbf{k}  -  \textbf{k}^\prime)\,,\\
\langle \omega^i (\textbf{k}) \omega^{*j} (\textbf{k}^\prime) \rangle =& (2\pi)^3 \frac{1}{2} \bigg( \delta^{ij}-\frac{k^ik^j}{k^2} \bigg) P_{\omega \omega}(\textbf{k}) \delta ( \textbf{k}  -  \textbf{k}^\prime)\,,\\
\langle \bm{v} (\textbf{k}) \cdot \bm{v}^* (\textbf{k}^\prime) \rangle =& (2\pi)^3  P_{vv}(\textbf{k}) \delta ( \textbf{k}  -  \textbf{k}^\prime)\,,\\
\langle \delta (\textbf{k} ) \delta^* (\textbf{k}^\prime) \rangle =& (2\pi)^3 P_{\delta \delta}(k) \delta ( \textbf{k}  -  \textbf{k}^\prime)\,,
\end{align} 
where indices $i,j$ denote the components in the Fourier space coordinates and one can verify the velocity power spectrum satisfying the following relation,
\begin{equation}
 k^2 P_{vv} = P_{\theta \theta} + P_{\omega \omega}\,.
\end{equation}
In the linear perturbation theory, the continuity equation leads to $\theta=-\mathcal{H}f\delta$, where $\mathcal{H} =aH$ is the conformal Hubble parameter,  $a$ denotes the cosmic scale factor and $f$ is the linear growth rate in $\Lambda$CDM, defined by $f = d\ln D/d\ln a$, and $D$ is the linear density growth factor. In simulations, one can also consider the cross-spectrum of the velocity divergence $\theta$ with the overdensity $\delta$, i.e.,
\begin{equation}
\langle \delta (\textbf{k} ) \theta^* (\textbf{k}^\prime) \rangle = (2\pi)^3 P_{\delta \theta}(k) \delta ( \textbf{k}  -  \textbf{k}^\prime)\,,
\end{equation}

As proposed in~\citep{2019MNRAS.487..228B,2017PhRvD..95f3527C}, the velocity dispersion tensor is defined as the variance of the velocities of multi-streams at a given point, weighted by their respective local density on each stream, 

\begin{equation}
  \sigma_{ij}^2 (\textbf{x}) = \langle v_i (\textbf{x}) v_j (\textbf{x}) \rangle - \langle v_i (\textbf{x}) \rangle \langle v_j (\textbf{x}) \rangle\, 
\end{equation}
where the stream averaging is given by  
\begin{equation} \langle f(\textbf{x}) \rangle = \frac{ \sum_k \rho^{(k)} (\textbf{x}) f^{(k)} (\textbf{x}) }{ \sum_k \rho^{(k)} (\textbf{x})}\,,
\end{equation}
and hence, the velocity dispersion tensor becomes
\begin{equation}
 \sigma_{ij}^2 (\textbf{x}) = \frac{\sum_k \rho^{(k)}v_i^kv_j^{(k)}}{\sum_k \rho^{(k)}} -\frac{\sum_k\rho^{(k)}v_i^{(k)}}{\sum_k\rho^{(k)}}\frac{\sum_k\rho^{(k)}v_j^{(k)}}{\sum_k\rho^{(k)}},
\end{equation}
where the index $k$ runs over all streams that contain point $\textbf{x}$, and
the fields $\rho^{(k)}$ and $v^{(k)}$ are linearly interpolated to the evaluation point using the values of the vertices on the tetrahedron. The velocity dispersion power spectrum is defined by
\begin{equation} \langle \sigma^2 (\textbf{k}) \sigma^2 (\textbf{k}^\prime)^* \rangle = (2\pi)^3 P_{\sigma^2 \sigma^2}(\textbf{k}) \delta ( \textbf{k}  -  \textbf{k}^\prime)\,,
\end{equation}
with $\sigma^2 \equiv {\rm Tr}(\sigma^2_{ij}) $.

The density and velocity fields are estimated by DTFE (Delaunay tessellation method) public code~\citep{2011arXiv1105.0370C}, setting the number of meshes as 1000. We use six-point difference method to calculate the velocity divergence and vorticity fields with periodic boundary, and their power spectra are computed by using the massively parallel toolkit, Nbodykit~\citep{2018AJ....156..160H}.

As known, a measurement of the volume-weighted velocity statistics from $N$-body simulations is challenging, because of an unphysical sampling artifact in estimating the velocity field from the particles to the regular grids. Many methods~\citep{10.1093/mnras/279.2.693,Zheng_2013,Koda_2014} have been proposed for the velocity field estimation in cosmological $N$-body simulations, and recently the Kriging method~\citep{Yu_2015,Yu_2017} and the Delaunay Tessellation Field estimator (DTFE)~\citep{2011arXiv1105.0370C} are promising to provide high accuracy in the velocity field reconstruction. In this study, we use the DTFE to construct both density and velocity fields, with setting the number of meshes as 1000, which provides a natural multidimensional (linear) interpolation grids for estimating them from the particle positions to regular grids. And, numerically, we apply a six-point difference scheme to estimate the velocity divergence and vorticity fields with periodic boundary conditions. As known, this finite-difference scheme in the estimation of velocity divergence and vorticity could become problematic due to the multi-valued nature of the velocity field. In spite of this, the power spectra of these fields are not strongly affected by the choice of finite-difference scheme, and we find the estimates are robust when the nonlinear scales are well resolved. In addition, contrary to other velocity components, due to numerical errors during the estimation, spurious vorticity --- especially at high redshifts where the shell-crossing occurs rarely --- would arise to some extent, and hence, in this study we focus on vorticity results at present-day, $z=0$, which are expected to be reliable.

\subsection{Simulations}

Our $N$-body simulations were performed using the code $\texttt{CUBE}$ described in~\cite{2020ApJS..250...21I}, which is an improved version of ${\rm CUBEP^3M}$~\citep{2013MNRAS.436..540H}. The code we used develops a new method to resolve neutrino perturbations by decomposing the neutrino phase space into shells of constant speed and evolving those shells using hydrodynamic equations.
It is well known that, the particle-based method can naturally capture complete nonlinear neutrino clustering, whereas, due to a large thermal motion of neutrinos, this method inevitably suffers from Poisson noise on small scales. Such Poisson noise could be reduced by increasing the number of simulated neutrino particles, which, however, will lead to tremendous storage and computational overhead. In contrast to the particle-based method, various other methods have been proposed to implant massive neutrinos into the standard N-body simulations, while aiming to avoid this Poisson noise problem, e.g., the grid-based method~\citep{2009JCAP...05..002B}, the linear response approximation~\citep{2013MNRAS.428.3375A,2018MNRAS.481.1486B}, the particle- and grid-based hybrid approach~\citep{2010JCAP...01..021B} and the fluid technique~\citep{2016JCAP...11..015B,2017PhRvD..95f3535I,2020ApJS..250...21I,2021arXiv211015867Y}. In fact, the grid-based and the linear response approaches cannot accurately resolve the non-linear neutrino structure formation on small scales, which however can be alleviated by the hybrid approaches (detailed in the above-mentioned literature). Recently, based on low-resolution and high-resolution neutrino N-body simulations, the fluid technique is promising to highly reduce the Poisson noise contamination and could resolve non-linear neutrino clustering evolution accurately. Furthermore, from various simulations the maximum suppression from massive neutrinos on matter power spectrum typically occurs at around $k = 1 h/{\rm Mpc}$ (as we will see in Fig.~\ref{wuzhi}), and we find such suppression from the fluid-technique based $\texttt{CUBE}$ simulation does deviate slightly from the particle-based one. More importantly, in this study, all our quantitative analyses on the sensitivity of density and velocity fields on the neutrino mass sum (in Sect.~\ref{s5}) are based on the degree of discrepancy between massive- and massless- neutrino power spectra, all from the $\texttt{CUBE}$ code of the fluid technique, rather than comparing with the linear theory predictions or the particle-based results. Therefore, our conclusions are self-consistent and reliable.

The cosmological parameters adopted for the simulations are compatible with the Planck 2018 results~\citep{2020A&A...641A...6P} for a spatially flat universe, which, specifically, are $\{\Omega_m, \Omega_b, h, 10^9A_s, n_s\} = \{0.32, 0.05, 0.67, 2.215, 0.96 \}$, and the initial conditions are generated by perturbing a regular lattice of dark matter particles according to the Zel'dovich approximation at $z = 100$ using the Boltzmann solver CAMB~\citep{2000ApJ...538..473L}. The constructed initial velocity field is vorticity-free from this approximation.

In our simulations, baryon physics is not included, so that, in the following, we will define $\Omega_c \equiv \Omega_{\rm cdm}+\Omega_b$ as the density corresponding to the sum over cold dark matter and baryon densities, and, for simplicity, it will be referred to as ``CDM'', denoted by the subscript ``$c$''. Throughout the paper we plot CDM power spectra and ratios as observed galaxies directly trace the cold dark matter and baryon fluids but not the total matter.

We run one CDM-only simulation ($M_\nu=0.0$) and three CDM+neutrino simulations with neutrino masses $M_{\nu}=$ 0.1, 0.2 and 0.4 eV respectively, where $M_{\nu}=\sum m_i$, and $m_i$ is the mass of each type of neutrino. The mass sum corresponds to the energy densities of massive neutrinos in terms of
\begin{equation}\label{eq:nu}
\Omega_\nu = \frac{M_\nu}{93.14 h^2\rm eV}\,.
\end{equation}
Consequently, the corresponding values of $\Omega_{\nu}$ and $\Omega_c$ vary with changing $M_\nu$ accordingly, with keeping $\Omega_m$ and the primordial amplitude $A_s$ fixed, where $\Omega_m = \Omega_c+\Omega_\nu$. In the case of massless neutrinos, the simulation with $M_\nu=0.0$ eV is thus equivalent to the $\Lambda$CDM case.

For the CDM+neutrino simulations, only a single neutrino species is massive, corresponding to a normal mass hierarchy. Here, our simulations only employ the particle-mesh algorithm with $n_c=256^3$ coarse grids and $n_f=4 \times n_c=1024^3$ coarse grids for Fast Fourier Transforms (FFTs). To avoid the contaminations of the cosmic variance in our data analysis, all the simulations use the same initial Gaussian random noise field to generate the initial conditions of $1024^3$ CDM particles at redshift $z=100$, in a periodic cubic box of width $L = 600~{\rm Mpc}/h$. In CDM+neutrino simulations, the neutrinos are added at redshift $z=10$, and are solved by three fluids with different grid resolutions of $n_{\nu}^3=(n_f/2)^3=512^3$, $n_{\nu}^3=(n_f/4)^3=256^3$ and $n_{\nu}^3=(n_f/8)^3=128^3$. The Nyquist frequency of the simulations is ($k_N = \pi N^{1/3}/L_{\rm box} \approx 4.19~h/\rm Mpc$). For the investigation of the divergence and vorticity fields estimated from the finite-difference scheme, throughout the paper we only concern velocity-related power spectra in the range of $k \lesssim 2~h/$Mpc to avoid numerical errors and make results reliable.



\section{Impacts of massive neutrinos on CDM power spectrum}\label{s3}
\begin{figure*}
	\centering
	\includegraphics[width=0.75\textwidth]{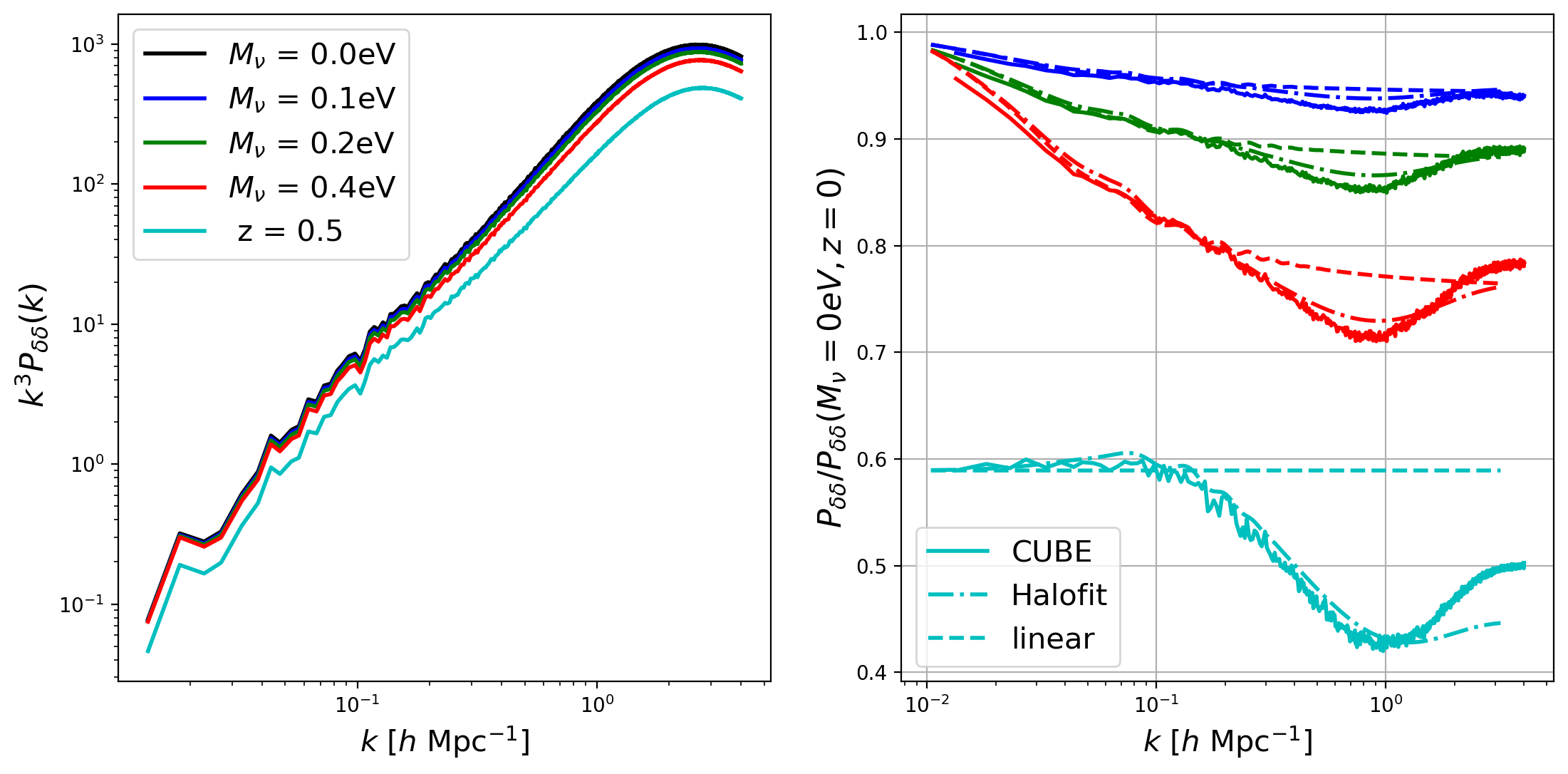}
        \includegraphics[width=0.45\textwidth]{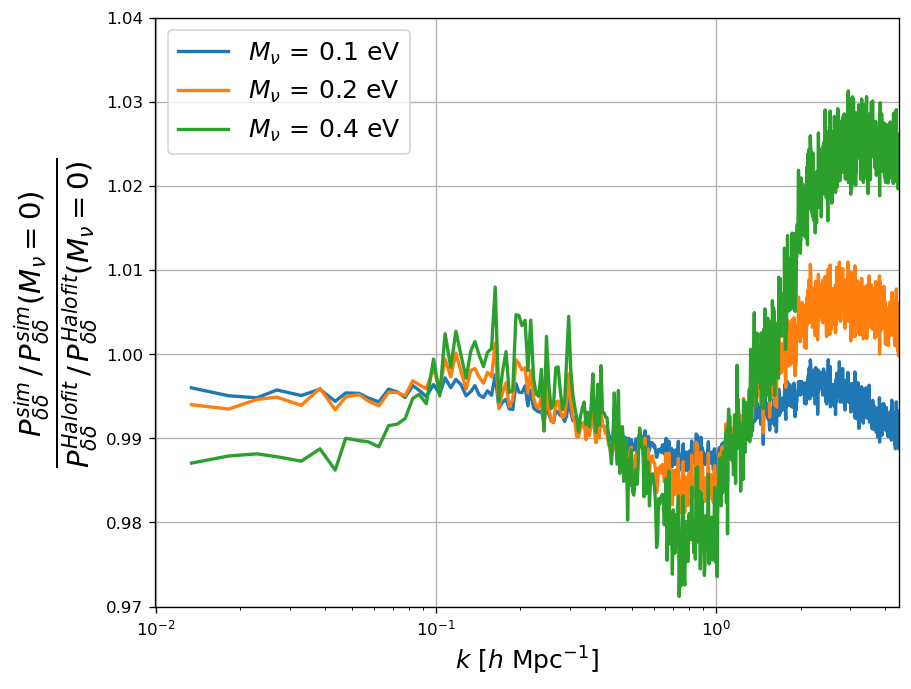}
	\caption{Comparison with CDM power spectra from different neutrino masses with respective to a $\Lambda$CDM one.  Upper left panel: the power spectra for various neutrino masses at $z=0$, and for masssless neutrino (i.e., the $\Lambda$CDM case) at $z=0.5$, from the $\texttt{CUBE}$ simulations. Upper right panel: corresponding ratios $P_c(k,z=0)/P^{\Lambda\rm CDM}_c(k,z=0)$ for $M_\nu=0.1$ eV (blue), $M_\nu=0.2$ eV (green) and $M_\nu=0.4$ eV (red), respectively.  Cyan curves show the ratios in the massless scenario, $P^{\Lambda\rm CDM}_c(k,z=0.5)/P^{\Lambda\rm CDM}_c(k,z=0)$. The ratios predicated from the linear theory (dotted) and the Halofit (dash) are also presented for comparison. As seen, a well-known spoon-like suppression is confirmed by our results. Lower panel: ratios of ratios, $\frac{P^{\rm sim}_{\delta\delta}/P^{\rm sim}_{\delta\delta}(M_\nu=0)}{P^{\rm Halofit}_{\delta\delta}/P^{\rm Halofit}_{\delta\delta}(M_\nu=0)}$, for the three netrino masses, in which the effects due to different mass resolutions are expected to be highly canceled out. Compared with the Halofit predictions, the relative deviations for $M_\nu$ =0.1 and 0.2 eV basically fluctuate in the range of about $1\%$ level, and it can reach to $2$--$2.5\%$ for the extreme case of $M_\nu=0.4$ eV in the strongly nonlinear regime when $k\gtrsim 2 h{\rm/Mpc}$. Thus, our simulation results agree well with the Halofit predictions at $k< 1 h{\rm/Mpc}$, at least at the $2\%$ level of accuracy.}
  	\label{wuzhi}
\end{figure*}

It is well known that, massive neutrinos can slow down the growth of perturbations of CDM and baryons on both linear and non-linear scales, as their large thermal velocities prevent them from gravitational clustering. Therefore, the suppression from massive neutrinos relative to the linear CDM power spectrum in $\Lambda$CDM is approximately given by~\cite{Castorina_2015,2012arXiv1212.6154L}:
\begin{equation}\label{eq:pcc}
  \frac{P_c^{\nu{\rm CDM},\rm lin} (k)} {P_c^{\Lambda{\rm CDM},\rm lin}(k)}\approx 1- 6 f_\nu\,,
\end{equation}
where $f_\nu$  represents the neutrino fraction defined as $f_\nu \equiv \Omega_\nu/\Omega_m$.

Contrary to the linear theory that predicts a constant suppression, many pioneering $N$-body simulations have observed a spoon-like feature in the massive-to-massless matter power spectrum ratio, $P_c(k)/P_c(k,M_\nu=0)$, where the suppression first increases up and then turns around, decreasing gradually to below even the linear-theory suppression at large $k$. From our simulations, we do confirm the characteristics of the spoon-like suppression and find that the maximum suppression occurs at $k\sim 1~h/\rm Mpc$ and become of order  $\Delta P_c/P_c\simeq 10.1f_\nu$, 10.1$f_\nu$, 9.7$f_\nu$ for $M_\nu= 0.1$, $0.2$, $0.4$ eV, respectively.

\cite{2020JCAP...11..062H} gives an analytical explanation of the origin of the spoon-like feature in the power spectrum at nonlinear regimes in terms of the standard halo model of large-scale structure. The spoon feature essentially originates in the transition from the two-halo power spectrum to the one-halo power spectrum. Specifically, the two-halo term is suppressed by free-streaming neutrinos and the sensitivity to $f_\nu$ increases with $k$, while the one-halo term is affected by neutrinos falling into CDM halos and its sensitivity decreases with $k$.

However, as pointed out by~\cite{2014JCAP...12..053M}, we should notice that, this spoon-like shape is not a unique feature of massive neutrino cosmologies (although the massive neutrinos can deepen the suppression), such shape appears also when the power-spectrum ratio between two identical $\Lambda$CDM models but with different values of $\sigma_8$, since the one-halo term changes with different $\sigma_8$-dependent halo-mass functions accordingly. To support this argument, we have computed the ratio of power spectra in absence of massive neutrinos, $P^{\Lambda\rm CDM}_c(k,z=0.5)/P^{\Lambda\rm CDM}_c(k,z=0)$, shown in the upper right panel of Fig.~\ref{wuzhi}. One can see that, due to a slight change in $\sigma_8$ which varies from 0.955 for $z=0$ to 0.708 for $z=0.5$, the spoon shape is also clearly present even without massive neutrinos, but the suppression can now be exactly reproduced by the Halofit model.

\begin{figure}
	\centering
	\includegraphics[width=0.45\textwidth]{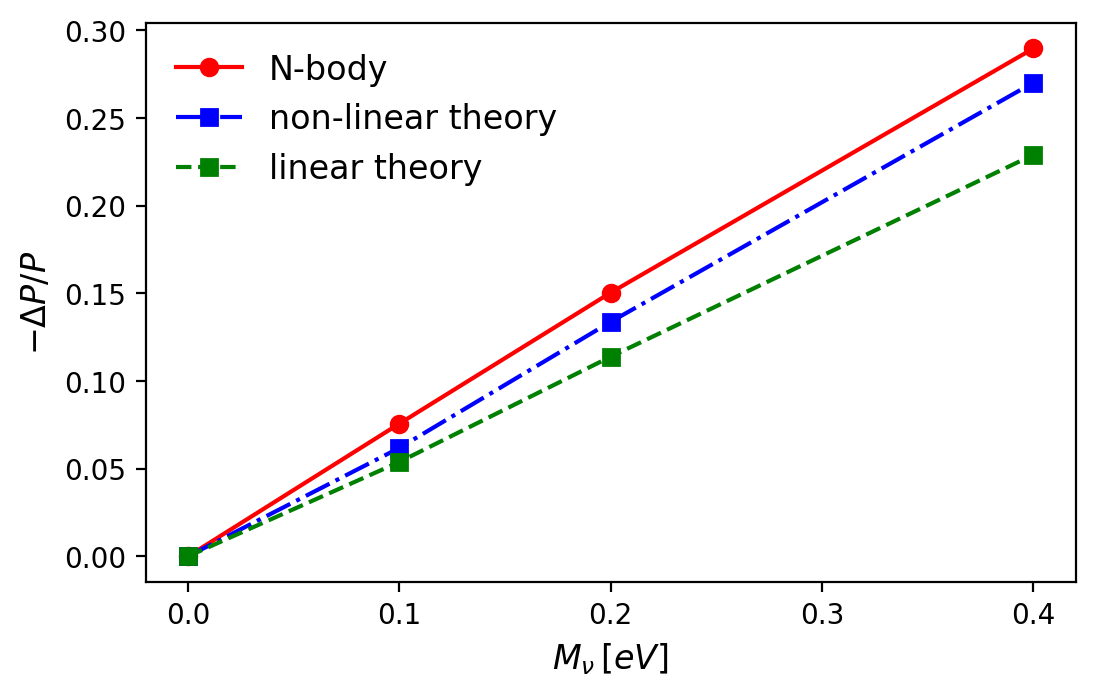}
	\caption{Maximum suppresion in CDM power spectrum as a function of the neutrino mass. The maximum suppression occurs at $k\simeq 1~h/\rm Mpc$. The suppresion from the $\texttt{CUBE}$ simulations (solid-red) from $M_\nu= 0.4$ eV is stronger than that from the linear theory (dash-green), and that from the non-linear theory based on the Halofit model HMcode-2020~\citep{2021MNRAS.502.1401M} (dash-blue). The simulation result shows that the suppression strength almost linearly depends on the total neutrino mass.}
  	\label{kmax}
\end{figure}

\section{Impacts of massive neutrinos on peculiar velocity}\label{s4}
In this section, we show the effects of massive neutrinos on velocity, velocity divergence, vorticity, and velocity dispersion, from the previously described $\texttt{CUBE}$ simulations. We first give a visual impression of the spatial distribution of various velocity-related fields. We then measure the corresponding power spectra of those fields to quantitatively determine the massive-neutrino induced impacts.

\begin{figure*}
	\centering
	\includegraphics[width=0.9\textwidth]{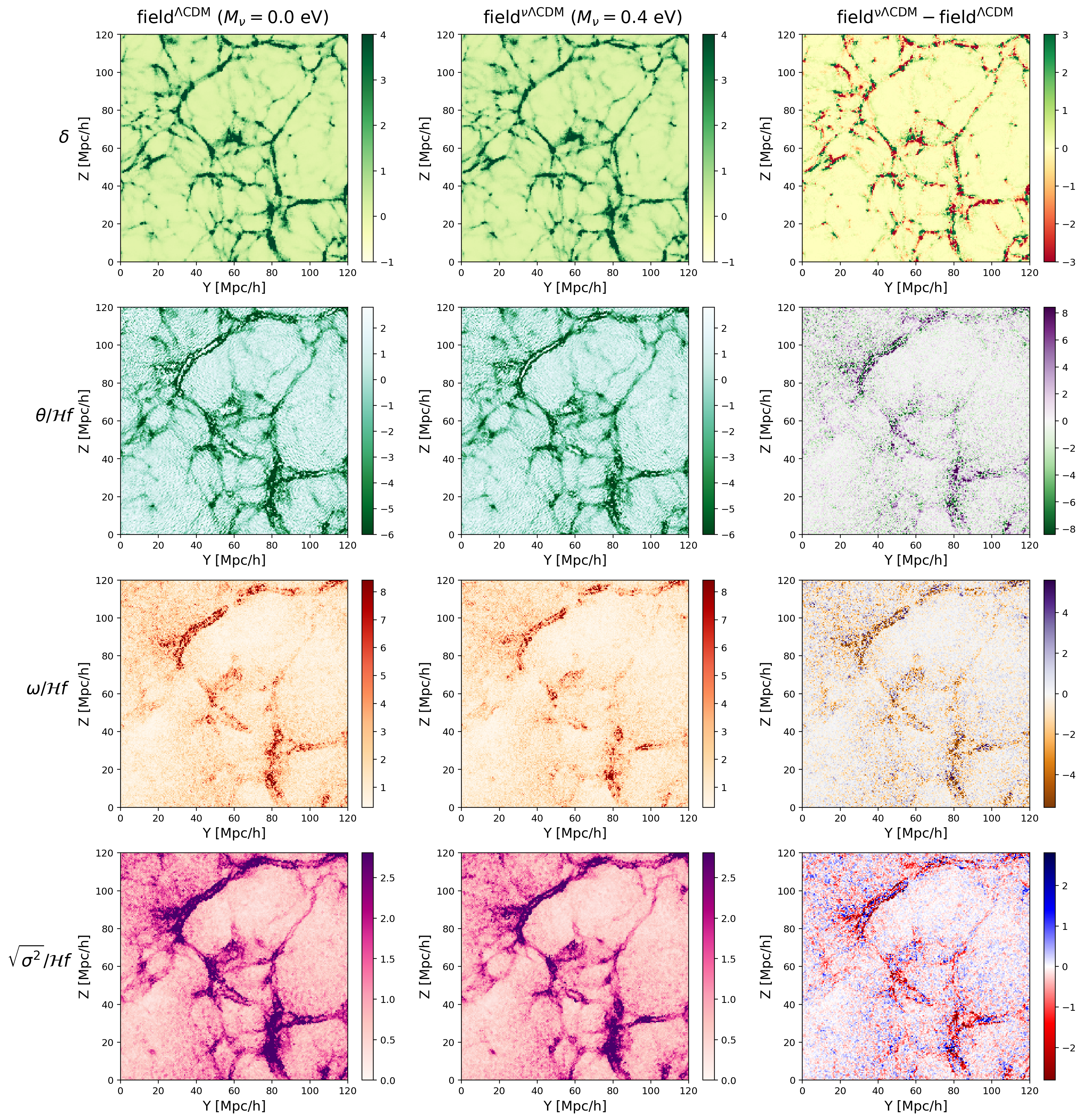}
	\caption{Slices of the cosmic overdensity (top row) and volume-weighted velocity fields: divergence (2nd row), vorticity magnitude (3rd row), dispersion magnitude(bottom row), from the $\texttt{CUBE}$ simulations estimated from the DTFE method,  with a size of 120 ${\rm Mpc}/h$ and each grid cell of 600 ${\rm kpc}/h$ at $z=0$. The left and middle columns are taken from simulations without neutrinos ($M_\nu=0$ eV) and with massive neutrinos ($M_\nu=0.4$ eV). The right column shows the contrast relative to the case of massless neutrinos to highlight the impacts of massive neutrinos.}
  	\label{redu}
\end{figure*}

\subsection{Snapshots in CUBE simulations}

We provide our primary results on impacts of massive neutrinos on the peculiar velocity, which have not been studied as deeply as the density field. The understanding of impacts of neutrinos on the volume-weighted (as opposed to density weighted) velocity field could be very important for neutrino cosmology, since the volume-weighted statistics of large scale peculiar velocity is independent of galaxy bias and will provide a clean observable to accurately measure the neutrino masses from future observations.

In Fig.~\ref{redu} from top to bottom, we show slices of the density, the velocity divergence, the vorticity and the velocity dispersion fields at $z$ = 0, respectively, extracted from a comoving box size of $600~{\rm Mpc}/h$ with $1024^3$ CDM particels in the $\texttt{CUBE}$ simulations. Each grid cell in our simulations is 600 ${\rm kpc}/h$. The left column is from a $\Lambda$CDM simulation without neutrinos (i.e., setting $M_\nu=0$ eV), and the middle one corresponds to a $\nu\Lambda$CDM simulation with $M_\nu=0.4$ eV, and the right one show their difference by $\rm {field}^{\nu\Lambda\rm CDM} - {\rm field}^{\Lambda\rm CDM}$ for a clear comparison.

For the density field (top row of Fig.~\ref{redu}), when neutrinos become massive, they will suppress the growth of density perturbations at small scales, leading to the appearance of lower density and more diffuse halos and filaments (dark green regions) in comparison with the massless case, as expected from the power spectrum results discussed above. Also, it is evident that  massive neutrinos make the voids less underdense. Note that, the overdensities in dark matter halos can be much larger than the upper scale limit we have chosen, $\delta_{\rm max} =4$, whereas the high-density regions are much smaller relative to the scale of the plot and the low-density structures would become invisible if increasing the $\delta_{\rm max}$ significantly.

We use the dimensionless quantities, $\theta/(\mathcal{H}f)$ and $|\bm\omega|/(\mathcal{H}f)$, to characterize the divergence ($\theta = \nabla \cdot \bm{v}$) and the vorticity ($\bm{\omega} = \nabla \times \bm{v}$) components in plots, which is convenient because in the linear theory, this normalized divergence is directly related to the dimensionless overdensity in terms of $\theta = -\mathcal{H}f\delta$.  For the divergence field (the 2nd row of Fig.~\ref{redu}), one can see that the features are remarkably similar to those in the overdensity field, as predicted from the linear theory. Moreover, the regions with negative values of the divergence field essentially corresponds to the overdensed regions ($\delta >0$). This  pattern is just indicative of dark matter particles inflowing into the high-density regions. In the center of the high density regions, we observe lower in-flow velocity compared to the surrounding space, and even out-flow in some cases. This result is consistent with the findings in ~\cite{2009PhRvD..80d3504P,2015MNRAS.454.3920H}. Furthermore, due to lower $\delta$ from the massive-neutrino induced suppression, one can see that, the massive-neutrino induced convergence field around density peaks, in absolute value ($|\theta|$), would become relatively smaller compared with the $\Lambda$CDM case. In addition, the divergence fields for the both cases has more substructures at small scales than that in the density field and become more extended in spatial distribution, which are due to the fact that velocity divergence field would become more fluctuating than the linear-theory prediction when it well enter into the nonlinear regime. Such non-linear evolution of the peculiar velocity field would lead to amplitudes of the divergence appearing randomly fluctuations at very small scales (e.g., see the bottom-left corner at the right panel of the plot about $\theta$). However, small-scale velocity convergence field may not provide more information on constraining neutrino masses.

The spatial distributions of the normalized vorticity magnitude, $|\bm\omega|/(\mathcal{H}f)$, is present in the 3rd row of Fig.~\ref{redu}. In linear perturbation theory, any existing vorticity in the linear regime of structure formation will only rapidly decay due to the universe expansion, so that the generation of vorticity implies nonlinear physics occurring somewhere. In principle, the emergence of anisotropic stress in the Euler equation can yield vorticity. In cosmological context, at small scales, gravitational collapse and associated nonlinear structure formation, where shell-crossing is occurring, will lead to emergence of nontrivial stress tensor, and consequently create vorticity in CDM distribution. In a standard cosmology, there are no physical processes to generate large-scale fluctuations with a coherence length larger than 1 ${\rm Mpc}/h$, a typical cluster scale, in vorticity, as seen in Fig.~\ref{redu} of $|\bm\omega|/(\mathcal{H}f)$, so that the vorticity power spectrum is expected to be considerably small in low-$k$ regime (as we will discuss in Sec.~\ref{psvel}). As seen, the vorticity field is mainly concentrated on collapsing regions, and thus it is tightly coupled to the local density. Physically, \cite{2014ApJ...793...58W,2015MNRAS.454.3920H} have shown that, a DTFE-estimated vorticity field can be approximately given by $\left<\bm\omega\right> \sim \left<\nabla\log\rho \times (\bm{v}-\left<\bm v\right>)\right>$, which implies that massive neutrinos will highly suppress the vorticity magnitude through their suppression effects in both density and velocity. Theoretically, the vorticity evolution is sourced by three terms~\citep{2009PhRvD..80d3504P}, $\bm\omega\theta$, $\bm \omega^2$ and $\bm \pi_{\omega} = \nabla \times \bm\pi$, where $\bm\pi$ is related to the velocity dispersion $\pi_i = (\nabla_j \rho \sigma^2_{ij} ) / \rho$. At large scale where the $\bm \omega$ is still perturbative (i.e. $k\lesssim 0.3~h/{\rm Mpc}$ for $P_{\omega\omega}\lesssim 1$), the vorticity is mainly sourced by the velocity dispersion. Consequently, in this regime, we can see that the neutrino suppression of $P_{\omega\omega,\nu}/P_{\omega\omega}$ ({\it lower-left} panel of Fig.~\ref{omegasigma}) is consistent with the suppression of the velocity dispersion $P_{\omega \omega, \nu}/P_{\omega \omega}$.

The bottom row in Fig.~\ref{redu} illustrate the normalized amplitude of the velocity dispersion, $\sqrt{\sigma^2}/(\mathcal{H}f)$, which is a dimensionless quantity and corresponds to sum of the dispersion along its main axes. The anisotropic nature of gravitational collapse leads to the emergence of an anisotropic velocity dispersion after shell-crossing in multistreaming regions. We observe the spatial distribution of velocity dispersion that is positively correlated with the density field, similar to the results found in~\cite{2019MNRAS.487..228B} that $|\sigma|^2 \propto (1+\delta)^\alpha$ with $\alpha \sim 0.5$ -- $1$. Compared to the $\Lambda$CDM case, the presence of massive neutrinos would lead to relatively lower amplitudes in the dense regions as this positive correlation.

\subsection{Impacts of massive neutrinos on velocity power spectra}\label{psvel}

\begin{figure*}
	\centering
	\includegraphics[width=0.75\textwidth]{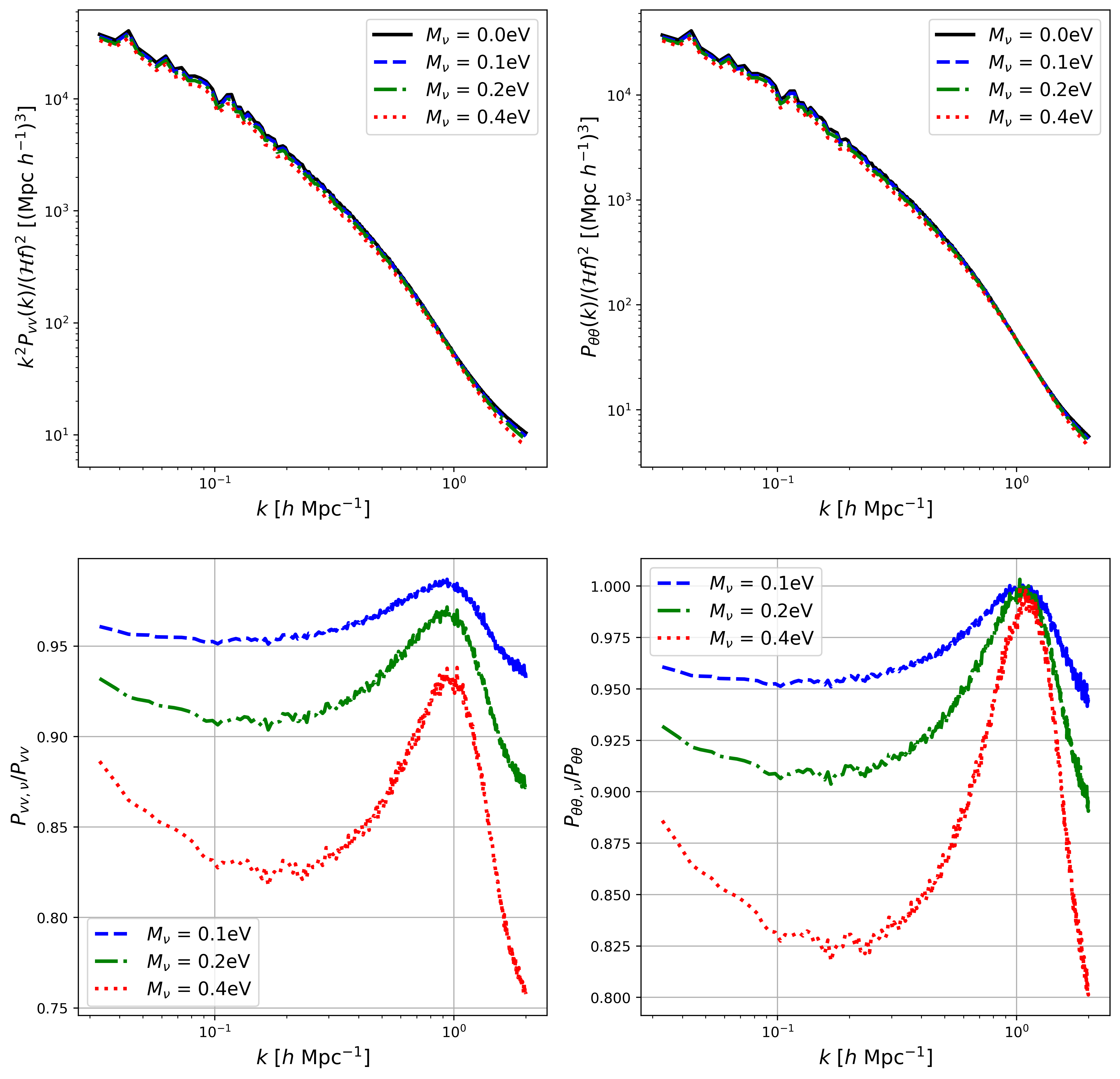}
	\caption{Top panels: DTFE-derived velocity-velocity (left) , velocity divergenc-velocity divergence (right) power spectra for different neutrino masses at $z=0$ from the $\texttt{CUBE}$ simulations. Bottom panels: the corresponding ratios between the massive neutrino-induced power spectra and the $\Lambda$CDM power spectra of $M_{\nu} = 0$.}
  	\label{vtheta}
\end{figure*}

To accurately quantify the difference between the simulations with and without massive neutrinos, we measure the auto-power spectra of various components of the velocity field, including velocity power spectrum $P_{vv}$, divergence spectrum $P_{\theta\theta}$, vorticity spectrum $P_{\omega\omega}$ and dispersion spectrum $P_{\sigma^2 \sigma^2}$, and a cross-power spectrum between the velocity divergence and the matter density contrast $P_{\delta \theta}$.

\begin{figure*}
  \centering
	\includegraphics[width=0.75\textwidth]{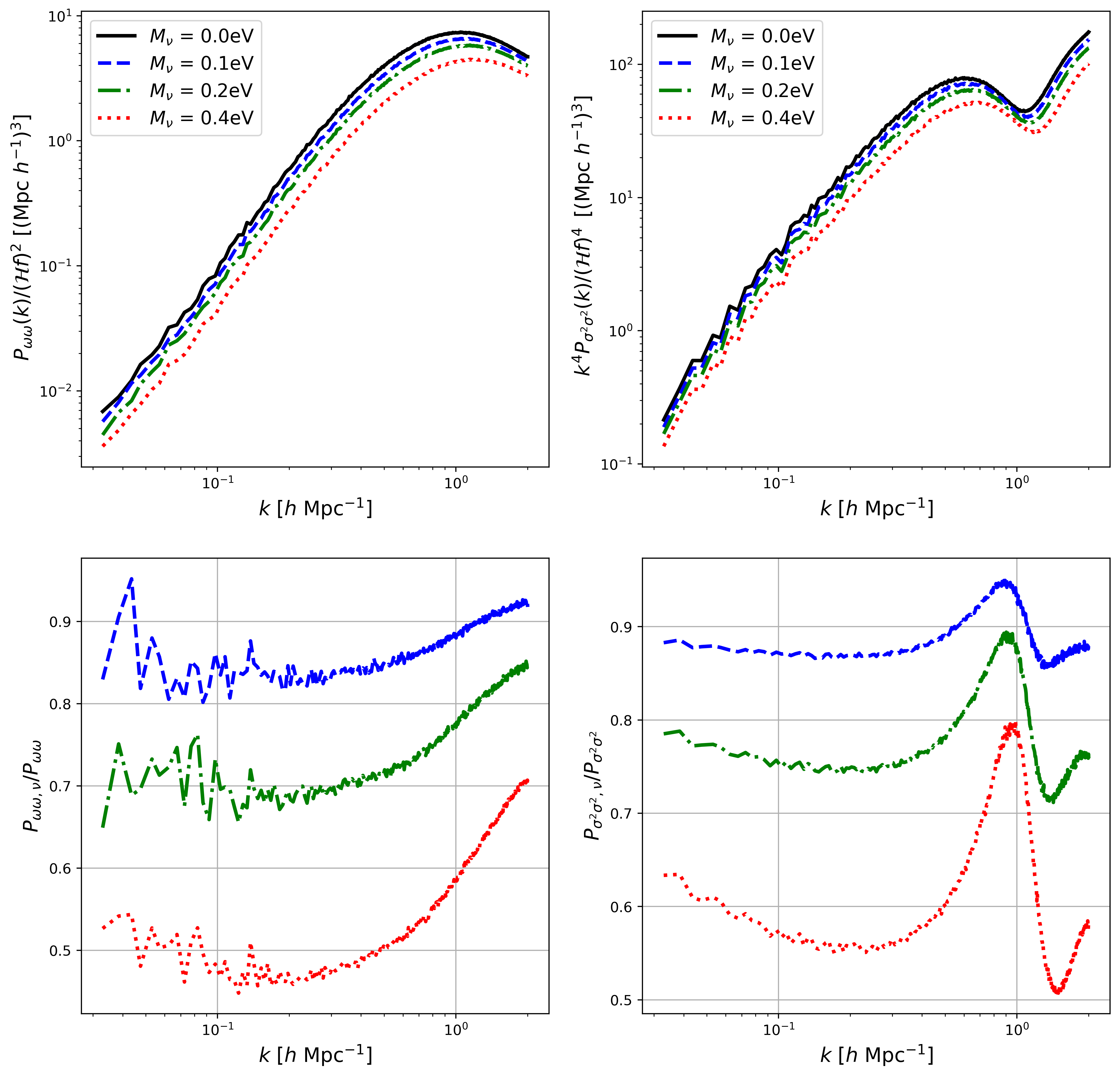}
	\caption{Top panels: the power spectra of vorticity $P_{\omega\omega}$ (left) and velocity dispersion $P_{\sigma^2\sigma^2}$ (right), for different neutrino masses at $z=0$. Bottom panels: corresponding ratios of power spectra, $P_{\omega\omega}$ (left) and $P_{\sigma^2\sigma^2}$ (right), with respect to their respective massless-neutrino results.}
   \label{omegasigma}
\end{figure*}

Fig.~\ref{vtheta} shows the measured auto-power spectra of velocity magnitude and velocity convergence for different neutrino masses at $z=0$, and the ratios to their respective massless neutrino-induced spectra. We notice that, the amplitudes and shapes for both $P_{vv}(k)$ and $P_{\theta\theta}(k)$ are highly similar, due to the fact that the curl component $P_{\omega\omega}$ of the velocity (discussed later in Fig.~\ref{omegasigma}) is significantly small compared with the divergence one such that $k^2P_{vv}(k) \approx P_{\theta\theta}$. The quantity of $k^3P_{\theta\theta}$ goes up rapidly with increase $k$, leading strong fluctuations at small scales, as seen in Fig~\ref{redu}. We also observe that, remarkable suppression on the power spectra, $P_{vv}(k)$ and $P_{\theta\theta}(k)$, caused by the massive neutrinos, and the extent of suppression increases with an increased $M_\nu$, by from a few percents to tens of percents. However, a notable feature in the spectra is that, the suppression becomes less effective for $P_{vv}$ and almost disappears for $P_{\theta\theta}$ (suppressed in $\sim1\%$ level), when $k$ approaching $k=1~h/{\rm Mpc}$. We suspect the main reason for this is that, this specific scale is around the boundary scale of the largest halos, where the velocity transfer from in-fall mode to stochastic one. In structure formation, the halo scale is essentially determined by the total matter fraction $\Omega_m$ (fixed in the $\texttt{CUBE}$ simulations) and is not sensitive to the neutrino fraction $f_\nu$ (equivalent to $M_\nu$). Thus, the neutrino impact is mild on $P_{vv}$ and $P_{\theta\theta}$. Additionally, the suppression feature at $k>1~h/\rm Mpc$ seems to indicate that the neutrino can highly affect the velocity properties below the typical cluster halo scale. A possible explanation is that the substructures are less in the massive neutrino case due to the delay in the structure formation.  However, the small scale is highly anisotropic and numerical artifacts can easily appear in the velocity interpolation, so that we will leave it for a future investigation.


\begin{figure}
	\centering
	\includegraphics[width=0.45\textwidth]{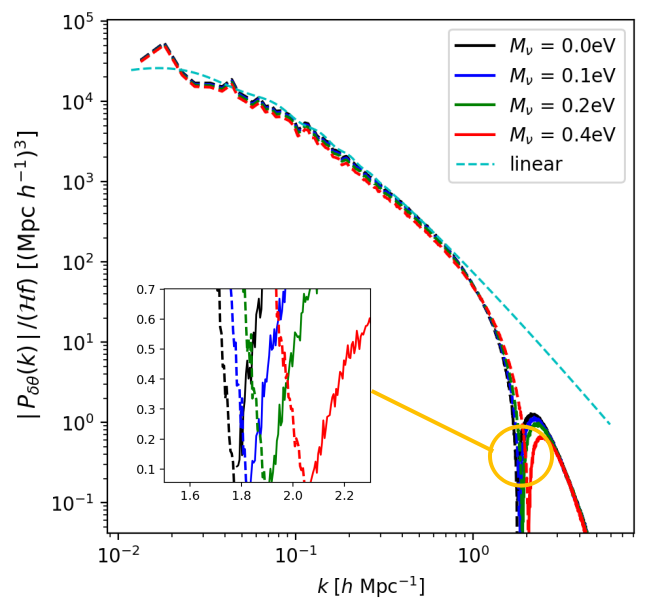}
	\caption{Cross-spectrum of overdensity and velocity divergence for different neutrino masses at $z=0$. The dash lines represent the negative values of the cross spectra, and the solid lines represent the positive cross. The cross spectra change sign from negative to positive at small scales around $k\sim 2~h/{\rm Mpc}$ due to the shell-crossing. The inset shows the enlarged view of the region with the sign changed in $P_{\theta\delta}$, highlighting the effects from the massive neutrinos. The linear-theory prediction (cyan dash) is  also shown from comparison, calculated by $P_{\delta\theta}= -\mathcal{H}f P_{\delta\delta}$, implying a constantly negative overdensity-divergence cross power spectrum over all scales.}
  	\label{guanlian}
\end{figure}

In Fig.~\ref{vtheta}, at large scales, there is a strong anti-correlation between the density contrast and velocity divergence as expected from the linearized continuity equation, $\theta = - \mathcal{H}f\delta$. At small scales, interestingly, the cross spectrum $P_{\theta\delta}$ changes sign at $k\sim 2~h/{\rm Mpc}$, and then the divergence and density contrast become positively correlated, which manifests the relevant scale of shell-crossing~\citep{2018JCAP...09..006J}. This is because, after shell-crossing, the formation of structures yields outward flows from high-density regions, leading to the positive correlations. Moreover, since the massive neutrinos can suppress the power spectra of both density and velocity divergence fields, thus we see a lower amplitude of $P_{\theta\delta}$ as increasing $M_\nu$. In addition,  the position of transition from negative to positive value of $P_{\theta\delta}$ is slightly shifted towards high $k$, from about 1.8 -- 2.0 $h/\rm Mpc$ by varying $M_\nu$ from 0 -- 0.4 eV. This shift indicates a decrease in the typical scale of shell-crossing regions, since massive neutrinos can slow down the structure formation and delay the time of non-linear collapse.

The vorticity power spectrum for the $\texttt{CUBE}$ simulation is shown in Fig.~\ref{omegasigma}, which can be well characterized by a power-law spectrum with spectral-index of $n_\omega$, i.e., $P_{\omega\omega}(k) \propto k^{n_\omega}$. We find, $n_\omega\approx (2.550 \pm 0.001)$ for $k\lesssim0.4~h/{\rm Mpc}$ and $n_\omega\to -1.5$ on small scales for $k\gtrsim 1~h/{\rm Mpc}$ by fitting this power-law fucntion to our simulation results, which are well consistent with the findings of ~\cite{Carrasco:2013mua,2015MNRAS.454.3920H,2018JCAP...09..006J}. In addition, these spectral indices are almost insensitive to $M_\nu$. Compared with the velocity divergence spectrum in Fig~\ref{vtheta}, the vorticity power spectrum peaks at $k\approx 1~h/{\rm Mpc}$ whereas the divergence spectrum exhibits a dip approximately at the same position. The peak position of  $P_{\omega \omega}(k)$ roughly corresponds to a cluster scale on which nonlinear structures become more common. Since the conservation of angular momentum will prevent particles from further falling and will lead to the particles rotating around a certain structure, and hence, while the small-scale nonlinear structures are forming, a fraction of the power in the divergence could be transferred into the vorticity. On the other hand, according to the relation, $k^2 P_{vv} = P_{\theta \theta} + P_{\omega \omega}$, although divergence component dominates the total velocity spectrum on large scales, yet the vorticity spectrum on small scales increases much faster than divergence one and eventually when $k\gtrsim 2~h/{\rm Mpc}$, gradually becoming the dominant component of the total velocity power spectrum.

We also find, in the presence of massive neutrinos, the vorticity power spectrum is highly suppressed by about $10\%-60\%$ for $M_\nu=0.1-0.4$ eV at $k\lesssim1~h/{\rm Mpc}$ and by about $8\%-30\%$ when $k\simeq 2~h/{\rm Mpc}$. These suppression effects are more pronounced than those in $P_{\delta\delta}$, $P_{\theta\theta}$ and $P_{vv}$ as well as $P_{\theta\delta}$. This is not surprising because the vorticity is very sensitive to nonlinear structure formation which can be slowed down by neutrino masses. Even though a small change in collapse formation at an earlier time will lead to a significant suppression in the vorticity spectrum at present-day, since the vorticity originates from a purely nonlinear process, generated by shell-crossing of the particles in the $\texttt{CUBE}$ simulations.

For the velocity dispersion, as expected, it would be strongly correlated with the large-scale density field, due to the fact that the shell-crossing mainly occurs in overdense regions. In Fig.~\ref{omegasigma}, we find the $P_{\sigma^2\sigma^2}(k)\propto k^{-1}$ at very large scales when $k\lesssim 0.05~h/{\rm Mpc}$, whereas at small scales it drops rapidly when $k\simeq  1~h/{\rm Mpc}$, implying a characteristic scale of the largest collapsed structures at present~\citep{2019MNRAS.487..228B}. As seen, the massive-neutrino induced suppression almost follows the same trend as that found in the density and various velocity-related spectra as discussed above, on average decreasing $P_{\sigma^2\sigma^2}$ by 10\%, 20\% and 40\% for $M_\nu=$ 0.10, 0.2 and 0.4 eV, respectively. The suppression effect becomes less important when $k$ approaching $1~h/{\rm Mpc}$, where the shell crossing of the particles occurs.          

\begin{figure*}
	\centering
	\includegraphics[width=0.45\textwidth]{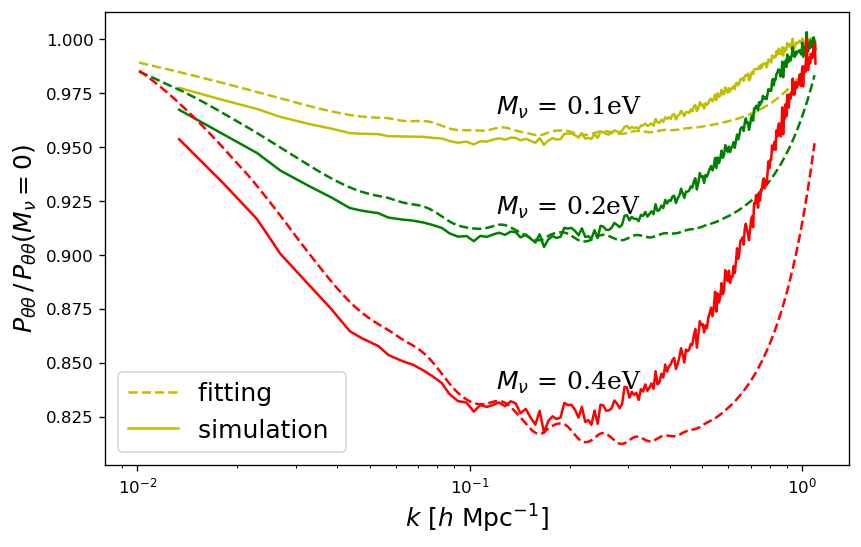}
        \includegraphics[width=0.45\textwidth]{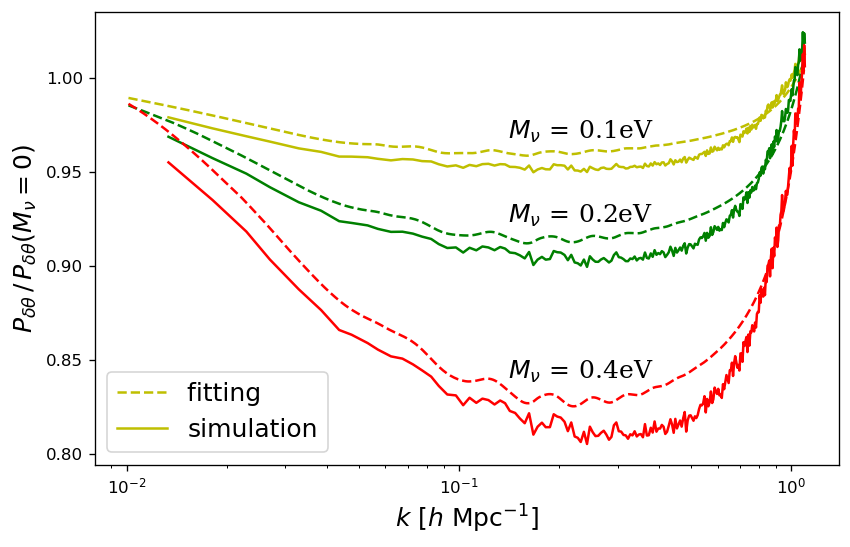}
	\caption{Comparison of the simulated velocity spectra  and the fitting formulae proposed in~\citep{2019A&A...622A.109B} (Eqs.~10 and 11 therein) for various neutrino masses at $z=0$. Left panel: ratios of the simulated massive-neutrino velocity divergence spectrum to the massless-neutrino one, $P^{\rm sim}_{\theta\theta}/P^{\rm sim}_{\theta\theta}(M_\nu=0~\rm eV)$ (solid) and $P^{\rm fit}_{\theta\theta}/P^{\rm fit}_{\theta\theta}(M_\nu=0~\rm eV)$ (dotted), are shown for $M_\nu= 0.1$ eV (yellow), 0.2 eV (green), 0.4 eV (red), respectively. Right panel: same as the left one, but for the cross-spectrum of overdensity and velocity divergence, $P_{\delta\theta}$.}
  	\label{fig:ratio}
\end{figure*}

~\cite{2019A&A...622A.109B} provides fitting formulae for massive neutrinos based on $\sigma_8$, describing the nonlinear corrections with $3$--$5\%$ accuracy on scales below $k\simeq 1 h/\rm Mpc$. In Fig.~\ref{fig:ratio}, we summarize the deviations of divergence and overdensity-divergence power spectra, $P_{\theta\theta}$ and $P_{\delta\theta}$, from the massless-neutrino spectra. As seen, the comparison between the simulation results and the fitting formula validates that our results are compatible with the fitting formulae at $1$--$5\%$ accuracy. One has to note that, the intrinsic accuracy of the fitting formulae is about $3$--$5\%$, so that our simulation results agree well with the fitting models.

\section{quantify the sensitivity to massive neutrinos}\label{s5}

\begin{figure*}
	\centering
	\includegraphics[width=0.65\textwidth]{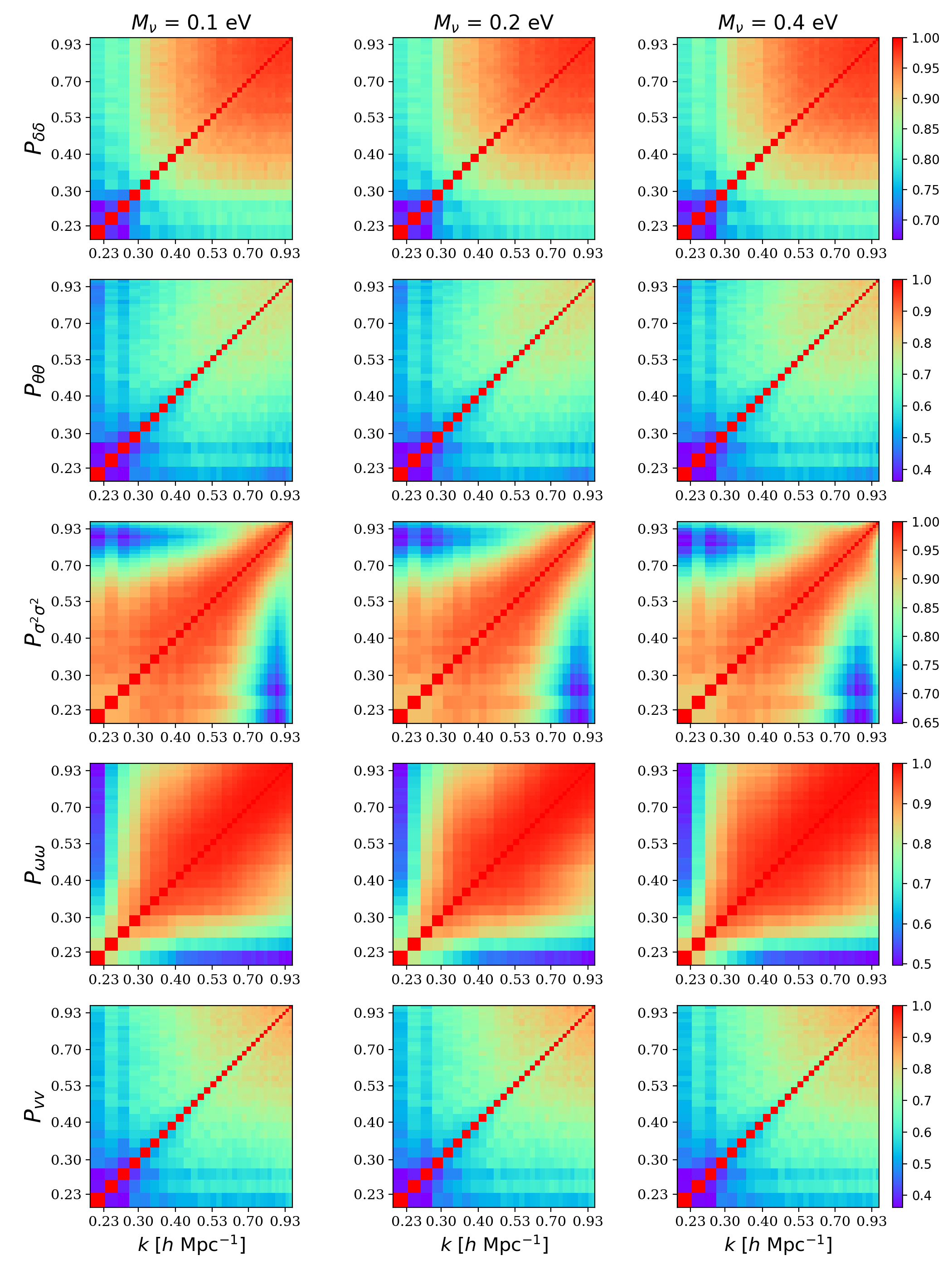}
	\caption{Normalized covariance matrices ($C_{ij}/\sqrt{C_{ii}C_{jj}}$) at $z=0$ for the power spectra of density, velocity divergence, vorticity, velocity dispersion, with $M_\nu=$ 0.1, 0.2, 0.4 eV, respectively, with 32 $k$-bins of bin-width $\Delta k = 0.025~h/$Mpc in the range of $k\in [0.207, 0.983]~h/{\rm Mpc}$.}
  	\label{cov_n}
\end{figure*}

\begin{figure}
	\centering
	\includegraphics[width=0.5\textwidth]{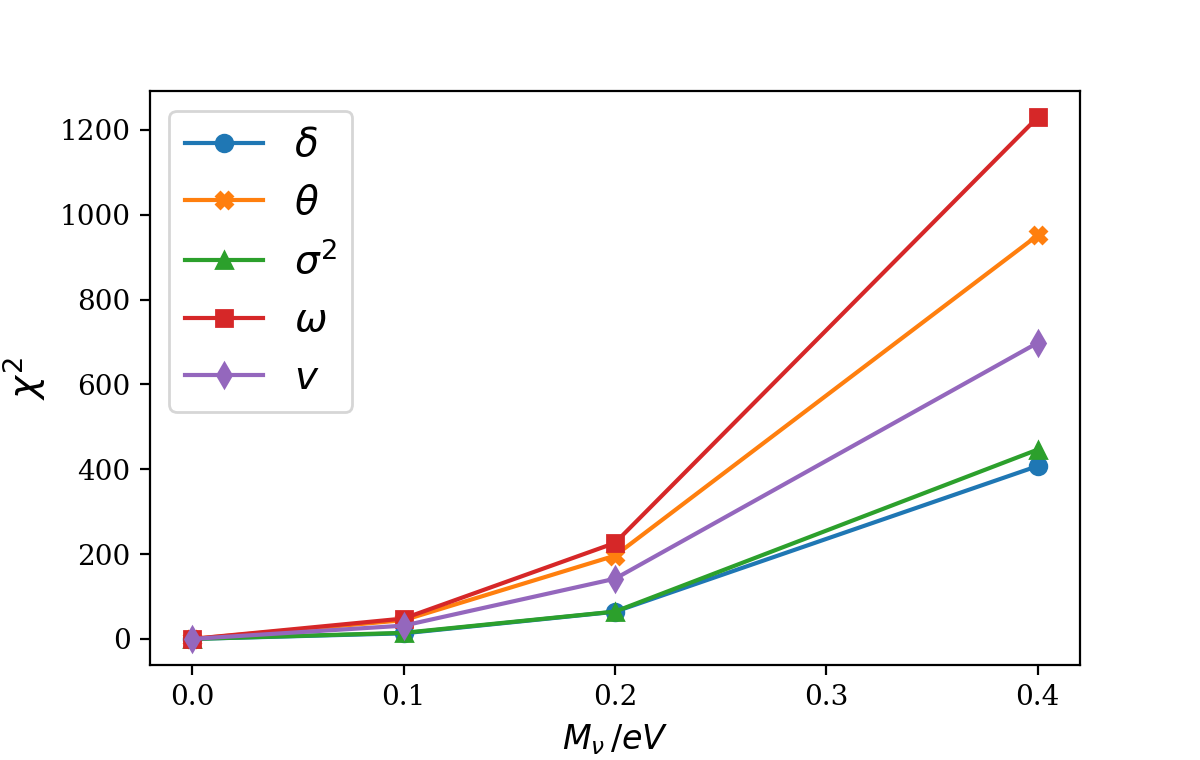}
	\caption{Sensitivity of massive neutrinos at $z=0$, as a function of $M_\nu$ for the power spectra of density, velocity magnitude, divergence and vorticity.  The calculations of $\chi^2$ are based on the 1000 jackknife sub-boxes with $k$-modes of $\Delta k=0.025~h/$Mpc in the interval $k\in[0.207,0.983]~h/{\rm Mpc}$, spanning over 32 $k$-bins, and hence with $32$ degrees of freedom.}
  	\label{kafang}
\end{figure}

\begin{table}
  \caption{$\chi^2$ and associated PTEs for the null hypothesis ($M_\nu=0$) for various power spectra and neutrino masses, with 32 degrees of freedom.  }
  \label{tab:chi}
  \begin{tabular}{c|c|c|c|c|c|c}
    \hline
    \multirow{2}{*}{field} &
      \multicolumn{2}{c|}{$ M_{\nu}=0.1$ eV} &
      \multicolumn{2}{c|}{$M_{\nu}=0.2$ eV} &
      \multicolumn{2}{c}{$M_{\nu}=0.4$ eV} \\
    & $\chi^2$ &  PTE & $\chi^2$ &  PTE & $\chi^2$ &  PTE \\
    \hline
    $P_{\delta\delta}$ & 13.17 & 0.998 & 64.03 & 0.0006  & 407.64 & $\simeq 0$ \\
    \hline
    $P_{vv}$ & 31.29 & 0.502 & 141.91 &  $\simeq 0$ & 698.17 & $\simeq 0$ \\
    \hline
    $P_{\theta\theta}$ & 44.10 & 0.075 & 196.32 &  $\simeq 0$ & 951.73 &  $\simeq 0$ \\
    \hline
    $P_{\omega\omega}$ & 48.08 & 0.033 & 226.31 & $\simeq 0$ & 1229.92 &  $\simeq 0$ \\
    \hline
    $P_{\sigma^2\sigma^2}$ & 14.29 & 0.997 & 64.42 & 0.0005 & 446.40 &  $\simeq 0$ \\
    \hline
  \end{tabular}
\end{table}

In order to quantitatively test the sensitivity of various power spectra to massive neutrinos, we use a standard chi-square approach by comparing massive-neutrinos induced spectra with massless-neutrinos (i.e., a $\Lambda$CDM model) induced ones. In other words, this quantity describes the measured power spectra for neutrino masses against the null hypothesis (i.e., $M_\nu=0$), which reads  
\begin{equation}\label{eq:chi}
  \chi^2_{\alpha}=  \left(\bm{p}_\alpha -\bm{p}_\alpha^{\Lambda\rm CDM}\right)^T \bm{C}_{\alpha}^{-1}\left(\bm{p}_\alpha-\bm{p}_\alpha^{\Lambda\rm CDM}\right)\,,
\end{equation}
where $\alpha$ refers to the power spectrum of a given field discussed above, $\alpha\in \{ P_{\delta\delta}, P_{\theta\theta}, P_{\omega\omega}, P_{vv}, P_{\sigma^2\sigma^2} \}$,  and  $\bm{p}$ and $\bm{p}^{\Lambda\rm CDM}$ represent the power spectrum data vectors for the ``massive-neutrino'' and ``massless-neutrino'' cases, respectively. The covariance matrix is estimated by applying jackknife methods~\citep{2021arXiv210907071M} to a single mock data with high accuracy and precision. The jackknife realisations are built by deleting only one of $n_{sv}$ sub-samples each time, and calculating the power spectrum/correlation function for the remaining data. Specifically, for a given $\alpha$, the covariance $\bm{C}$ is determined via the delete-one jackknife analysis, dividing the simulation box, $(600~{\rm Mpc}/h)^3$ with $1000^3$ cells in total, into $n_{sv}=1000$ sub-boxes with equal volume. As such, each jackknife box size contains $100^3$ cells.  The jackknife estimate of the covariance matrix reads
\begin{equation}\label{eq:cov}
C_{ij} = \frac{n_{sv}-1}{n_{sv}}\sum_{l=1}^{n_{sv}}\left(P^{[l]}(k_i)-\bar{P}(k_i)\right) \left(P^{[l]}(k_j)-\bar{P}(k_j)\right)\,,
\end{equation}  
where $P^{[l]}(k_i)$ denotes the power spectrum at the $i$-th $k$ bin calculated  from the $l$-th jackknife realization, and the mean estimate from $n_{sv}$ jackknife realizations is
\begin{equation}\label{eq:mean}
\bar{P}(k_i) = \frac{1}{n_{sv}}\sum_{l=1}^{n_{sv}} P^{[l]}(k_i)\,.
\end{equation}
In order to make our results robust and conservative, only the scales of $k<1~h/{\rm Mpc}$ are taken into account in the $\chi^2$ analysis. In practice, we compute the jackknife covariance matrix for a given field with 32 $k$-bins of bin-width $\Delta k = 0.025~h/\rm Mpc$ in the range of $k\in [0.207, 0.983]~h/{\rm Mpc}$.



The resulting covariance matrices for various power spectra at $z=0$ are illustrated in Fig.~\ref{cov_n} (we plot the normalized covariance, $C_{ij}/\sqrt{C_{ii}C_{jj}}$, for a clear illustration). Recently, ~\cite{Mohammed_2014, Carron_2015} have suggested the following ansatz to the covariance matrix,  $C_{ij} =P_{k_i}P_{k_j}\left(2\delta_{ij}/N_{k_i} + \sigma_{\rm min}P_{k_i}P_{k_j}\right)$, where $N_{k_i}$ is the number of Fourier modes associated to the $i$-th power spectrum bin. The first term corresponds to the Gaussian covariance and the second one approximates the shell-averaged trispectrum of the field~\citep{2014PhRvD..89h3519L}, so that the parameter $\sigma_{\rm min}$ describes non-Gaussian contribution from the structure formation, depending cosmological parameters and redshift. As seen from Fig.~\ref{cov_n}, since the massive neutrinos affect the structure formation as discussed above, they leave significant impacts on the correlation between different Fourier $k$-modes for density and velocity fields, resulting in the phenomenological parameter $\sigma_{\rm min}$ being sensitive to the value of $M_\nu$.

  The results of the sensitivity to neutrino masses for different power spectra are shown in Fig.~\ref{kafang}. One can see that, by varying $M_\nu$, the $\chi^2$ changes significantly for $P_{\omega\omega}$, from $\chi^2=48.08$ for $M_\nu=0.1$ eV to $1229.92$ for 0.4 eV, for 32 degrees of freedom. Using a statistical test, the probability-to-exceed (PTE) of the $\Lambda$CDM model (the null hypothesis) is $0.033$ for $M_\nu=0.1$ eV and rapidly approaches zero for heavier neutrino masses, indicating the data even for the small $M_\nu$ strongly incompatible with the null hypothesis. This is because that, the changes in $P_{\omega\omega}$ by increasing $M_{\nu}$ are significantly large (see Fig.~\ref{omegasigma}) and lead to the drastic discrepancy between the null hypothesis and the measured data. Moreover, we find that, one can not reject the null hypothesis from $P_{\delta\delta}$, $P_{\sigma^2\sigma^2}$ and $P_{vv}$  data when $M_\nu=0.1$ eV as the corresponding PTEs are not sufficiently small (PTE $<0.05$, a typical threshold for rejection), and the results are summarized in Tab.~\ref{tab:chi}. The density power spectrum is least sensitive to neutrino mass. In addition, using the velocity divergence data, $P_{\theta\theta}$, the PTE is about 0.075 for $M_\nu =0.1$ eV, which also indicates the measured spectrum is somewhat inconsistent with the null hypothesis at the confidence interval of $[68\%,95\%]$. Due to the large changes in $P_{\theta\theta}$ and small correlations between different $k$-bins in its covariance matrix, the $\chi^2$ increases more rapidly than other velocity components when increasing $M_\nu$. Thus, we can conclude that, even for $M_\nu=0.1$ eV, the vorticity spectrum $P_{\omega\omega}$, which is generated from non-linear structure formation, has the highest  sensitivity  to massive neutrinos and allow us to reject the $\Lambda$CDM model with high statistical confidence by using the $\chi^2$ test. From the tests, the velocity fields, especially for vorticity and divergence components, would be expected to be more promising than the density field for constraining the neutrino mass in current and future observations.

\section{conclusion and discussion}\label{s6}
In this paper, we have investigated the sensitivity of various physical observables to massive neutrinos, which is particularly important for accurately determining the sum of neutrino masses in view of present and upcoming LSS surveys. Using the $\texttt{CUBE}$ simulations that can accurately resolve neutrino perturbations by decomposing the Fermi-Dirac phase space into shells of constant speed and then evolving those shells using hydrodynamic equations. Specifically, we have quantitatively assessed the impacts of the total neutrino mass on the mass function, marked correlation functions, CDM power spectrum, and in particular, on the CDM peculiar velocity fields by measuring the power spectra of velocity magnitude, divergence and vorticity as well as dispersion, which provide more information on the influence of neutrinos on the nonlinear structure formation and evolution, and thus become more sensitive to the total neutrino mass.   

The key findings of this study can be summarized as follows: 
\begin{enumerate}


\item we confirm a spoon-like feature in the massive-to-massless matter power spectrum ratio, with the maximum suppression of power occurring at $k\sim 1~h/\rm Mpc$. The suppression from massive neutrinos exceeds the linear-theory and Halofit predictions, with of order $\Delta P_c/P_c\simeq 10.1f_\nu$, $10f_\nu$, $9.7f_\nu$ for $M_\nu= 0.1$, $0.2$, $0.4$ eV, respectively.

  

\item  in the presence of massive neutrinos, the velocity vorticity spectrum is more highly suppressed than other velocity and density components, which is because the massive neutrinos would substantially slow down the non-linear structure formation that can most significantly generate the vorticity than other velocity and density  components.


\item  using the chi-square approach by comparing the predictions between massive neutrinos and $\Lambda$CDM model, we tested the sensitivity of various power spectra to the total neutrino mass. We find that, the vorticity spectrum has the highest sensitivity to massive neutrinos and the divergence spectrum is second only to vorticity in sensitivity. The simulated vorticity spectrum even for $M_\nu=0.1$ eV is greatly incompatible with the null hypothesis of $\Lambda$CDM model.

\end{enumerate}

To our knowledge, observationally, accurate velocity measurements remain a challenge and are still fraught with problems, leading to errors that are difficult to eliminate. However, recently with new high-precision data, such as LSST~\citep{Ivezi__2019}, DESI~\citep{desicollaboration2016desi}, CSST~\citep{2019ApJ...883..203G}, Euclid~\citep{2011arXiv1110.3193L,2018LRR....21....2A}, and advanced techniques it is possible to reliably reconstruct the cosmic velocity fields, e.g., a deep learning technique to infer the non-linear velocity field from the dark matter density field~\citep{Wu_2021}, and a new Bayesian-based framework to infer the full three dimensional velocity field from observed distances and spectroscopic galaxies~\citep{Lavaux_2016}. In addition, the presence of vorticity would leave observable effects, e.g., on redshift space distortions and the alignment of halo spins~\citep{Laigle_2014}. Therefore, velocity field measurements in the near future are expected to provide a stronger constraining power on determining the mass of neutrinos, and we will leave this task for future study.

\section*{Acknowledgments}
We thank Jiaxin Han for useful discussions. This work is supported by the National Key R$\&$D Program of China (2018YFA0404504, 2018YFA0404601, 2020YFC2201600, 2020SKA0110401), National Science Foundation of China (11621303, 11653003, 11773021, 11890691, 11803094, 11903021), the Science and Technology Program of Guangzhou, China (No. 202002030360), the 111 project, the CAS Interdisciplinary Innovation Team (JCTD-2019-05), and the science research grants from the China Manned Space Project with No. CMS-CSST-2021-A03 and No. CMS-CSST-2021-B01.

\section*{Data Availability}
Data available on request.



\bibliographystyle{mnras}
\bibliography{cites} 





\bsp	
\label{lastpage}
\end{document}